\documentclass[12pt]{article}
\usepackage{amssymb,amsmath,esint,titlesec}
\titleformat*{\section}{\normalsize\bf}
\titleformat*{\subsection}{\small\bf}

\begin{document}


\begin{titlepage}

\setlength{\baselineskip}{18pt}

                               \vspace*{0mm}

                             \begin{center}

{\Huge\bf Toward a relative $\mathbf{q}$-entropy }

                                   \vspace{40mm}

              \Large\sf  NIKOLAOS \  KALOGEROPOULOS $^\dagger$\\

                            \vspace{3mm}

  \normalsize\sf  Department of Mathematics and Natural Sciences,\\
                           The American University of Iraq, Sulaimani,\\
                           Kirkuk Main Road, Sulaimani,\\
                              Kurdistan Region, Iraq. \\

                            \vspace{2mm}
                         
                                    \end{center}

                            \vspace{40mm}

                     \centerline{\normalsize\bf Abstract}
                     
                           \vspace{3mm}
                     
\normalsize\rm\setlength{\baselineskip}{18pt} 

We address the question and related controversy of the formulation of the $q$-entropy, 
and its relative entropy counterpart, for models described by continuous (non-discrete) sets of variables.  
We notice that an $L_p$ normalized functional proposed by  Lutwak-Yang-Zhang (LYZ), 
which is essentially a variation of a properly normalized relative R\'{e}nyi entropy up to a logarithm, 
has extremal properties that make it an attractive candidate which can be used to construct such a relative $q$-entropy.  
We comment on the extremizing probability distributions of this LYZ functional, its relation to 
the escort distributions, a generalized Fisher information and the corresponding Cram\'{e}r-Rao inequality.
We point out potential physical  implications of the LYZ entropic functional and of its extremal distributions.\\

                           \vfill

                                                                         
                             \vfill

\noindent\rule{8cm}{0.2mm}\\  
   \noindent   $^\dagger$ {\footnotesize\rm Electronic mail: \ \  \normalsize{nikos.physikos@gmail.com}}\\

\end{titlepage}
 

                                                                                \newpage                 

\rm\normalsize
\setlength{\baselineskip}{18pt}

\section{Introduction}

The $q$-entropy has been one of the simplest in functional form,  and probably the most studied, nonadditive entropic functional in the Stastitical Physics 
literature during the last thirty years. 
Initially it was introduced in Information Theory and Statistics, and after it was re-discovered by C. Tsallis \cite{T1}, it has been developed and brought 
to prominence ever since, for its potential applications to Statistical Physics, as well as to a variety of other fields not necessarily related to Physics at all \cite{T-book}. 
Despite the potentially wide range of its applicability, the dynamical foundations of the $q$-entropy remain quite obscure to this date.\\

 If one is convinced  that phase space hyperbolicity \cite{BP}, in the sense of Dynamical Systems \cite{KH}, lies at the heart of the success of the Boltzmann/Gibbs/Shannon (BGS) 
entropic functional, a natural question arises as to what is the corresponding dynamical origin of the $q$-entropy. An answer to this question has proved to be elusive, so far. 
An easier to address, and somewhat related, question has to do with the validity, or even existence, the $q$-entropy for systems described by variables taking values in a 
continuous set, as opposed to discrete sets of variables. 
Some controversy has arisen related to this issue during the last decade \cite{Abe1, Andresen, Abe2, BOT, LB, BL, PR, OB1, OA, OB2}.  We presented a possible resolution in 
\cite{CK} among the other recent proposals. The present paper can be considered as a  continuation along the line of these works 
\cite{Abe1, Andresen, Abe2, BOT, LB, BL, PR, OB1, OA, OB2, CK}. \\         
  
The present work relies very heavily on, and  provides some physical context and interpretation of, the results of \cite{LYZ0}. 
The results of \cite{LYZ0} can be seen in the wider context of landmark results of the dual $L_p-$ Brunn-Minkowski theory 
and the associated star-shaped bodies, which were introduced by E. Lutwak \cite{Lut1, Lut2, Lut3, Lut4}, and further developed jointly 
by E. Lutwak, D. Yang, G. Zhang \cite{LYZ1, LYZ2, LYZ3, LYZ4}, and collaborators \cite{HLYZ}.\\  
 
In Section 2, we present the $L_p$ relative entropic functional of Lutwak-Yang-Zhang (LYZ) and some of its properties. We claim this functional may 
contain a good candidate for the sought after  relative $q$-entropy for continuous systems and explain the reasons why. 
In Section 3, we present the generalized Gaussians which 
are the extremizing distributions of the LYZ functional  and some of their properties. In Section 4, we comment about the  Fisher information and its generalization 
and its potential physical implications.  In Section 5, we present a more general speculation about the dynamical 
foundations of the $q$-entropy via coarse-graining, the potential importannce of the duality among $L_p-$ Brunn-Minkowski theories in Convex Geometry for $q$-entropies, 
and their  conjectured possible invariance under changes of the non-extensive parameter \ $q$.\\


\section{Some relative entropy functionals}

To set up the notation, we consider the Boltzmann/Gibbs/Shannon (BGS) functional form for the entropy of discrete outcomes \ $i \in I$ \ with corresponding probabilities \ $p_i$ \ to be  
\begin{equation}
         \mathcal{S}_{BGS}[\{p_i\}] \ = \ - k_B \sum_{i\in I} \   p_i \log ( p_i ) 
\end{equation}
where \ $k_B$ \ is Botlzmann's constant.
The obvious extrapolation of the BGS entropy  to continous distributions is 
\begin{equation}
        \mathcal{S}_{BGS}[\rho ] \ = \ - k_B \int_X  \rho(x) \log(\rho(x)) \ dvol_X 
\end{equation}
where 
\begin{equation}
      \rho(x)\ = \ \frac{d\mu}{dvol_X}
\end{equation}
 is the Radon-Nikodym derivative of a chosen measure \ $\mu$, \  usually resulting after a process of coarse-graining, with respect to the volume measure \ $dvol_X$ \ of the 
``phase space" \ $X$, \ which is usually a Riemannian manifold or more generally, a metric measure space. In the latter case, we assume that \ $X$ \  is initially endowed with a 
reference measure \ $\nu$ \ and that the effective/coarse-grained measure \ $\mu$ \ is absolutely continuous with respect to\ $\nu$, \  so \ $\rho$ \ exists $\nu$-almost everywhere on \ $X$. \\
 
Given two  distributions \ $\rho_1, \rho_2: X \rightarrow [0,1]$, with respect to a reference measure \ $d\nu$, which may or may not be the volume of \ $X$, \  
their relative BGS entropy, or Kullback-Leibler divergence \ $\mathcal{D}_1 [\rho_1|\!|\rho_2]$ \ is defined \cite{CT} as 
\begin{equation}     
       \mathcal{D}_1 [\rho_1|\!| \rho_2] \ = \  \int_X \rho_1(x) \ \log \left( \frac{\rho_1(x)}{\rho_2(x)} \right) \ d\nu
\end{equation}
where we have arranged the units so that \ $k_B=1$ \ for brevity.
As is well-known, the Kullback-Leibler divergence \ $\mathcal{D}_1[\rho_1|\!|\rho_2]$, \ even though is not a metric, provides a way of 
measuring the discrepancy/difference between the densities \ $\rho_1$ \ and \ $\rho_2$. \ Interpreting the BGS entropy in a relative context, 
as a version of the Kullback-Leibler divergence, solves the issue of the lack of the diffeomorphism (reparametrization) invariance of \ $\mathcal{S}_{BGS}$, \
hence it allows the BGS entropy for continuous systems to potentially have physical meaning, from a formal viewpoint, something that is ultimately positively 
confirmed by its experimentally tested predictions.\\ 

With the above notation, the R\'{e}nyi entropy of order \ $\alpha\geq 0, \  \alpha \neq 1$ \  is defined for a discrete set of outcomes as
\begin{equation}       
          \widetilde{\mathcal{S}}_\alpha [\{p_i\ ] \ = \ \frac{1}{1-\alpha} \log \left( \sum_{i\in I} p_i^\alpha \right)
\end{equation}
Someone can readily check that 
\begin{equation}
         \lim_{\alpha\rightarrow 1} \widetilde{\mathcal{S}}_\alpha = \mathcal{S}_{BGS}
\end{equation}
For continuous probability distributions, with the above notation, the naive extension of the R\'{e}nyi entropy is 
\begin{equation}
           \widetilde{\mathcal{S}}_\alpha [\rho] \ = \ \frac{1}{1-\alpha} \log \left( \int_X [\rho(x)]^\alpha \ d\nu \right)    
\end{equation}
and the relative R\'{e}nyi entropy, in the continuous case, is 
\begin{equation}
      \widetilde{\mathcal{D}}_\alpha [\rho_1|\!|\rho_2] \ = \  \frac{1}{\alpha -1} \log \left( \int_X \rho_1(x) \left[\frac{\rho_1(x)}{\rho_2(x)}\right]^{\alpha -1} \ d\nu \right) 
\end{equation}
in analogy with the Kullback-Leibler divergence (4).\\

The $q$-entropy was initially introduced in \cite{HC, Vaj, Dar}, was more recently re-discovered in Statistical Physics and was proposed as an appropriate entropy for 
physical systems for which the BGS entropy may not be applicable in \cite{T1}. It is given, for discrete outcomes, by 
\begin{equation}
                \mathcal{S}_q [\{p_i \}] \ = \ \frac{1}{q-1} \left(1- \sum_{i\in I}p_i^q \right) 
\end{equation}
One can easily verify  that
\begin{equation}
                 \lim_{q\rightarrow 1} \mathcal{S}_q \ = \ \mathcal{S}_{BGS}
\end{equation}
For continous probabilities with density \ $\rho$ \ on a metric measure space \ $X$ \ with reference measure \ $d\nu$, \ the naive extension of (9)  is 
\begin{equation}
                 \mathcal{S}_q [\rho ] \ = \ \frac{1}{q-1}  \left(1- \int_X [\rho(x)]^q \  d\nu           \right)     
\end{equation}
Initially \cite{T1} it was assumed that the entropic/non-extensive parameter \ $q\in\mathbb{R}$, \ with a later proposal \cite{WW} for extending its domain to \ $q\in\mathbb{C}$. \  
A careful treatment of the possible values of \ $q$ \ that allow functional invertibility, as well as other desirable properties of the equilibrium distributions, called ``$q$-exponentials",
 resulting from a variational optimization of \ $\mathcal{S}_q$, \  which effectively encode the
 conjectured physical behaviors described by the $q$-entropic functional, \ was presented in \cite{OB10}. \\

In all of the above expressions the reference measure is taken to be the volume \ $dvol_X$ \ of \ $X$.\ The controversy regarding 
the suitability, or possibility of extending  the $q$-entropy for continuous systems \cite{Abe1, Andresen, Abe2, BOT, LB, BL, PR, OB1, OA, OB2}, is related 
to the relative version of (11) which, following (4),(8) is taken, by most authors working on this issue, to be  
\begin{equation}   
           \mathcal{D}_q [\rho_1|\!|\rho_2] \ = \ \frac{1}{q-1} \left(1 - \ \int_X \left[ \frac{\rho_1(x)}{\rho_2(x)} \right]^q \ dvol_X \right)
\end{equation}
A somewhat different approach to such a relative $q$-entropy was proposed in \cite{CK} based on generalized operations induced by \ $\mathcal{S}_q$. \
However, even the approach of \cite{CK} essentially relies on the functional form  (12).   \\ 

To move forward, we observe that in the R\'{e}nyi entropy \ $\widetilde{\mathcal{S}}_\alpha$ \ there are two parts of interest. One of them is its logarithmic behavior,
something that distinguishes it from the $q$-entropy and bears a strong similarity to the  functional form of the BGS entropy \ $\mathcal{S}_{BGS}$.  \ A second point
is the existence of the ``bias parameter" \ $\alpha$ \ in a power-law manner, strongly resembling the form of the $q$-entropy \ $\mathcal{S}_q$. \ We can disentangle 
these two behaviors by re-writing (7) as 
\begin{equation}
                   \widetilde{\mathcal{S}}_\alpha [\rho] \ = \ \log \left(\int_X [\rho (x)]^\alpha \ dvol_X  \right)^\frac{1}{1-\alpha}
\end{equation}
and (8) as       
\begin{equation}
                          \widetilde{\mathcal{D}}_\alpha [\rho_1|\!|\rho_2] \ = \ \log \left( \int_X \rho_1(x) \left[\frac{\rho_1(x)}{\rho_2(x)}\right]^{\alpha -1}\  dvol_X \right)^\frac{1}{\alpha -1} 
\end{equation}
The resemblance between the functional forms of (11), (13) is obvious. What separates these two functional forms is  the presence of the  logarithm in (13), and the additive unit in (11).
The latter may be  important for proper convexity and  for normalization purposes of the entropic functional, but it is far less important than the term next to it in (11).  
Other than that, the two parentheses in (11), (13) contain the same functional form.
For this reason, in the sequel, we will deal with this common term of the R\'{e}nyi and the $q$-entropies, in an attempt to find an expression for the most important part of the 
relative $q$-entropy, along the lines of a functional resembling the content of the parentheses of (14).  \\    

To proceed, we will recall  the definition of the $p$-norm of a function \ $f: X\rightarrow \mathbb{R}$ \ on the measure space \ $(X, vol_X)$ \ 
\begin{equation}
                        \Vert  f \Vert_p  \ = \  \left(  \int_X |f(x)|^p  \ dvol_X \right)^\frac{1}{p},   \ \ \   p\geq 1
\end{equation}
For \ $0<p<1$ \ this is only a quasi-norm since the triangle inequality is not obeyed, but this will not be an impediment in any of the future arguments.  Moreover, one defines
\begin{equation} 
          \Vert f \Vert_\infty \ = \ \mathrm{ess}\sup \{ |f(x)|, \  x \in X \}  
\end{equation}
where \ ess sup \ stands for the essential supremum of the function \ $f$. \ Given this standard notation, one can modify the definition inside the parentheses of (14), by normalizing 
\  $\rho_1, \ \rho_2$ in an \ $L_p$ \  rather than \ $L_2$ \ sense. \\

We will  confine ourselves to dealing with functions having as domain the set of reals \ $\mathbb{R}$,  \ for simplicity.
In  a general analytic context this is clearly a very strong simplification. However, when it comes to the convexity properties of interest in this work, and due to the localization technique
(``needle decomposition") \cite{PW, GroMil, KLS, Klartag} especially for a Riemannian space with a Ricci curvature uniformly bounded from below, considering a foliation by geodesics of the 
underlying space amounts to  essentially reducing convexity arguments from \ $X$  \ down to \ $\mathbb{R}$. \  Hence, considering aspects of the convex behavior of an entropy functional 
for functions defined over \ $\mathbb{R}$ \  instead of over \ $X$ \ may not be such a huge loss of generality as may appear at a first glance.\\         

Taking into account  the considerations of the above paragraph, one could propose that  instead of the argument of the logarithm of (14), one can consider   
\begin{equation}  
            \mathcal{N}_\lambda [\rho_1 \Vert \rho_2 ] \ = \ \left\{ \int_\mathbb{R} \ [\rho_1(x)]^\lambda \left[ \frac{\rho_1(x)}{\rho_2(x)} \right]^{1-\lambda} dx \right\}^\frac{1}{1-\lambda} 
                                                                  \frac{\Vert \rho_2\Vert_\lambda}{\left( \Vert \rho_1\Vert_\lambda\right)^\frac{1}{1-\lambda}}
\end{equation}
where the parameter \ $\lambda$ \  is employed, instead of using \ $\alpha$ \ as in (14) in order to align our notation with that of \cite{LYZ0}. \ 
A major difference between (14) and (17) is that \ $\rho_1(x)$ \ is linear  in the first factor of the integrand of (14), whereas it has a power-law dependence (raised in the power $\lambda$) in (17).    
Writing explicitly the norms of the functions, we get the LYZ functional for the relative entropy
\begin{equation}
            \mathcal{N}_\lambda [\rho_1\Vert \rho_2] \ = \ \frac{\left( \int_\mathbb{R} \rho_1(x) [\rho_2(x)]^{\lambda-1} dx \right)^\frac{1}{1-\lambda} 
                                           \left( \int_\mathbb{R} [\rho_2(x)]^\lambda dx \right)^\frac{1}{\lambda}}{\left(\int_\mathbb{R} [\rho_1(x)]^\lambda dx \right)^\frac{1}{\lambda(1-\lambda)}}, 
                                                \hspace{10mm} \lambda \neq 1
\end{equation}
and 
\begin{equation}
               \mathcal{N}_\lambda [\rho_1 \Vert \rho_2] \ = \ \exp(\mathcal{D}_1 [\rho_1\Vert \rho_2]), \hspace{10mm} \lambda = 1
\end{equation}
In (17), (18) the measure of integration is the Lebesgue measure of \ $\mathbb{R}$.\ One can define the \  $L_\lambda$  normalized relative R\'{e}nyi entropy by   
\begin{equation}
                           \mathcal{R}_\lambda [\rho_1\Vert \rho_2] \ = \ \log \mathcal{N}_\lambda [\rho_1\Vert \rho_2]
\end{equation}
It should be noticed that (20) is different from the argument of the logarithm of (14), so \ $\mathcal{R}_\lambda$  \ is not a simple variation of the conventionally defined 
relative R\'{e}nyi entropy \ $\mathcal{D}_\alpha$ \ (14), even up to a logarithm, but a totally different functional altogether.\\
  
In a similar spirit, the relative $q$-entropy can be inferred from \ $\mathcal{N}_\lambda $ \ (17) by omitting the overall exponent \ $1/(1-\lambda )$ \  in the first factor of (17) and normalizing 
the probability distributions in  an \ $L_q$ \ sense:
\begin{equation}
              \mathcal{T}_q [\rho_1\Vert \rho_2] \ = \  \frac{1}{q-1} \left\{ 1 -  \ \left( \int_\mathbb{R} \ [\rho_1(x)]^{2-q} \left[ \frac{\rho_1(x)}{\rho_2(x)} \right]^{q-1} dx \right)
                                                                  \frac{(\Vert \rho_2\Vert_q)^{q-1}}{ \Vert \rho_1\Vert_q}   \right\}
\end{equation}
which, in turn, gives
\begin{equation}
                            \mathcal{T}_q [\rho_1\Vert \rho_2] \ = \  \frac{1}{q-1} \left\{ 1- \frac{\left( \int_\mathbb{R}   \rho_1(x) [\rho_2(x)]^{1-q} dx \right)
                                           \left( \int_\mathbb{R} [\rho_2(x)]^q dx \right)^\frac{q-1}{q}}{\left(\int_\mathbb{R} [\rho_1(x)]^q dx \right)^\frac{1}{q}}                                       
           \right\}
\end{equation}
What we have done  in order to formulate (17), hence (20), and in the same spirit (21), (22) is to effectively renormalize the initial probability 
distributions \ $\rho_1, \ \rho_2$ \  in an \ $L_q$ \ sense, to the effective probabilities
\begin{equation}
                                      \overline{\rho}_i \ = \ \frac{\rho_i}{\Vert \rho_i \Vert_q}, \hspace{10mm} i=1,2
\end{equation}  
This is an occasion where the effective probability distributions \ $\overline{\rho}$ \ are ``escort distributions" \cite{T-book}, whose appearence is rather mysterious, on dynamical grounds 
at least, and  somewhat controversial even today \cite{BOB}. \\

One can recover the discrete form of the $q$-entropy (9), from (21) as follows: as a first step, consider the ``reference probability" distribution \ $\rho_2(x)$ \ to be the uniform one on 
the compact subset of \ $\mathbb{R}$ \ we are dealing with. If the support of such probability distribution is all of \ $\mathbb{R}$ \ then one has to implement a regularization procedure 
by confining themselves to probabilities having compact support (``putting the system in a box") and then taking a weak limit, as is frequently done in Quantum Physics, for instance.  This way 
(21) reduces to (11). As a second step, choose a discrete subset of \ $\mathbb{R}$ \ as the support of the probability distribution \ $\rho_1(x)$,  \ 
which now become essentially Dirac delta functions, up to normalization, and one recovers (9). It is evident, as is true most of the times, that the transition from the discrete to the continuous case 
and vice versa is not unique, but that this process  involves some judicious choices usually dictated by additional physical input and consistency requirements in taking the appropriate limits.  \\ 
     
The message that someone should take from the convex geometric considerations in this work 
seems to be: we do really have to use the renormalized ``escort distributions" rather than the ``usual" probability distributions, if we want our results to be more ``natural" 
and compatible with Convex Geometric considerations and also from the viewpoint of Information Theory. We believe that a major theoretical challenge is to find the dynamical 
reasons for the appearence of these ``escort distributions" starting from the Lagrangian or the Hamiltonian, ``microscopic", description of the systems of many degrees 
of freedom under consideration. We will consider a local, covariant, geometric quantity that may describe such a dynamical behavior in a near-future work \cite{NK8}. \\     


\section{Generalized Gaussians as entropy maximizers}

Given that the functional (17), (18) is the core of  equations (20) and (21), we now turn our attention to describing its extremal distributions and some key inequalities that it satisfies
following \cite{LYZ0}. To begin with, one defines the $p$-th moment of the probability \ $\rho: \mathbb{R} \rightarrow [0,1]$ \ as
\begin{equation}
              \mathfrak{m}_p [\rho] \ = \ \int_\mathbb{R} |x|^p \rho(x) \  dx,    \hspace{10mm} p\in (0, +\infty )
\end{equation}
as long as this integral exists. The $p$-th deviation \ $\sigma_p$, for \ $p\in [0,+\infty]$, \ is defined as   
\begin{equation}
             \mathfrak{s}_p [\rho ] \ = \ \left\{ 
               \begin{array}{ll}
                     \left( \mathfrak{m}_p[\rho] \right)^\frac{1}{p}        & \mathrm{if} \  \ p\in (0, +\infty) \\
                                                                                                         &               \\
                     \exp \left( \int_\mathbb{R} \rho(x) \log|x| \ dx\right) & \mathrm{if}  \  \  p=0\\
                                                                                                                      &                            \\
                     \mathrm{ess}\sup\{|x|: f(x)>0 \}                                              & \mathrm{if}  \   \ p=+\infty                                          
               \end{array}
                          \right.
\end{equation}
under the assumption that the above expressions exist and are finite. Given the symbol 
\begin{equation} 
              x_+ \ = \ \max\{x, 0\}, \hspace{10mm} x\in\mathbb{R}
\end{equation}
recalling the definition of Euler's Gamma function
\begin{equation}
               \Gamma (x) \ = \ \int_0^\infty z^{x-1}e^{-z} \ dz      
\end{equation}
and given the Beta function \ $B(x,y)$ \ 
\begin{equation}
                   B(x,y) \ = \ \int_0^1 z^{x-1}(1-z)^{y-1} \  dz 
\end{equation}
for \ $x>0, \ y>0$, \ one readily finds a relation between the Gamma and the Beta functions 
\begin{equation}               
                B(x,y) \ = \ \frac{\Gamma(x) \ \Gamma(y)}{\Gamma(x+y)}
\end{equation}
With this notation, the definitions of the generalized Gaussians \ $\mathcal{G}: \mathbb{R} \rightarrow [0, +\infty)$ \  
in an \ $L_p$ \ sense, for \ $p\in [0, +\infty]$ \ and for \ $a>1-p$, \ are as follows: 
\begin{equation}
    \mathcal{G}(x) \ = \ \left\{
             \begin{array}{ll}
                  c_{p,a} \left(1+(1-a)|x|^p\right)_+ ^\frac{1}{a-1} & \mathrm{if} \ \ a\neq 1\\
                                                                                                       &     \\
                  c_{p,1} \  \exp (-|x|^p)                                                 &  \mathrm{if} \ \ a=1 
             \end{array}  
                        \right.                   
\end{equation}
for \ $p\in (0, +\infty)$, \ with the normalization constant \ $c_{p,a}$ \  straightforwardly calculated to be
\begin{equation}
      c_{p,a} \ = \ \left\{
                  \begin{array}{ll}
                      \frac{p(a-1)^\frac{1}{p}}{2 B(\frac{1}{p}, \frac{a}{a-1})}  &  \mathrm{if} \ \ a>1\\  
                                                                                                                       &                         \\
                      \frac{p}{2\Gamma (\frac{1}{p})}                                           &  \mathrm{if} \ \ a=1\\ 
                                                                                                                           &                       \\
                      \frac{p(1-a)^\frac{1}{p}}{2 B(\frac{1}{p}, \frac{1}{1-a}-\frac{1}{p})} & \mathrm{if} \ \ a<1 
                 \end{array}
                           \right.
\end{equation}
For \ $p=0$ \ and \ $a>1$, \ the definition is 
\begin{equation}
             \mathcal{G}(x) \ = \ c_{0,a} (-\log |x|)_+ ^\frac{1}{a-1}
\end{equation}
for almost all \ $x\in\mathbb{R}$, \ with 
\begin{equation} 
             c_{0,a} \ = \ \frac{1}{2 \Gamma\left(\frac{a}{a-1}\right)}
\end{equation}
For \ $p=+\infty$ \  and \ $a >0$ 
\begin{equation}
      \mathcal{G}(x) \ = \ \frac{1}{2}, \hspace{10mm} -1\leq x \leq 1
\end{equation}
with \ $\mathcal{G}(x)=0$ \ everywhere else on \ $\mathbb{R}$, \ and  where we assume, for
notational consistency in the sequel that \ $c_{\infty, a} = \frac{1}{2}$. \ Moreover, we consider the re-scaled generalized Gaussians
\begin{equation}  
   \mathcal{G}_t (x) \ = \ \frac{1}{t} \  \mathcal{G}\left(\frac{x}{t}\right),  \hspace{10mm} t>0
\end{equation} 
A physical significance of such generalized Gaussians was established when it was proved in \cite{Baren}
that they are the self-similar solutions of the porous medium equation. The relation between the porous
medium equation and the $q$-entropy has been advocated by many authors such as \cite{PP, K, S, FD2, MMPL, LMMP, CN, NCR, T-book, SCN, RNC}, 
In \cite{NK9} we tried to explore  its possible significance for the dynamical underpinnings of the $q$-entropy. We also brought forth 
the significance of the developments that were initiated by the viewpoint of  by \cite{Otto} for systems described by the $q$-entropy, 
and subsequent developments  in the theory of metric measure spaces \cite{Villani, Ambrosio}.\\ 

At this point someone can define the ``absolute" analogue of the relative entropic functional (17), by
\begin{equation}
                           \mathcal{N}_\lambda [\rho] \ = \ \left\{
                                           \begin{array}{ll} 
                                                                       \left( \int_\mathbb{R} [\rho(x)]^\lambda \ dx \right)^\frac{1}{1-\lambda} & \mathrm{if} \ \ \lambda \neq 1\\
                                                                                                                                                                                               &                                  \\ 
                                                                        \exp \left( -\int_\mathbb{R} \rho(x) \log\rho(x) \ dx  \right)                        & \mathrm{if} \ \ \lambda = 1 
                                            \end{array}
                                                      \right.
\end{equation}
which, of course, is nothing else than the exponentials of the Renyi entropy (7) and of the Boltzmann/Gibbs/Shannon entropy (2) respectively. 
For the generalized Gaussians (30), (32), (34)  straightforward calculations give for their $p$-th deviation: for \ $0<p<+\infty$ \ and for \ $a>\frac{1}{1+p}$ \
\begin{equation}
        \mathfrak{s}_p [\mathcal{G}] \ =\ \frac{1}{(a-1+ap)^\frac{1}{p}}
\end{equation}
For \ $p=0$ \ and for \ $a>1$, \ one gets
\begin{equation}
          \mathfrak{s}_0 [\mathcal{G}] \ = \ \exp \left( - \frac{a}{a-1} \right)
\end{equation}
and for \ $p=+\infty$
\begin{equation}
                   \mathfrak{s}_\infty [\mathcal{G}] \ = \ 1 
\end{equation}
The absolute entropies (36) for these generalized Gaussians are, for \ $p\in (0,+\infty)$ \ and for \ $a>\frac{1}{1+p}$
\begin{equation}
              \mathcal{N}_a [\mathcal{G}] \ = \ \left\{
                          \begin{array}{ll}
                                   \frac{1}{c_{p,a}} (\frac{ap}{a-1+ap})^\frac{1}{1-a} & \mathrm{if} \ \ a\neq 1\\
                                                                                                                                          &                        \\
                                   \frac{1}{c_p,1} e^\frac{1}{p}                                                      & \mathrm{if} \ \ a=1
                           \end{array}
                                   \right.
\end{equation}
For \ $p=0$ \ and \ $a>1$ \ one gets
\begin{equation}
                \mathcal{N}_a [\mathcal{G}] \ = \ \frac{1}{c_{0,a}} \left(\frac{a}{a-1} \right)^\frac{1}{1-a}
\end{equation}
and finally, for \  $p=+\infty$ \ and \ $a>0$
\begin{equation}
                     \mathcal{N}_a [\mathcal{G}] \ = \ 2
\end{equation}

All of the above notation and results set the stage for the following statement. One of the ways that characterizes the ordinary Gaussians
and connects it with the BGS entropy, is the following extremizing property: among all the probability distributions with a given, finite, second moment,
the Gaussian is the unique probability distribution that maximizes the BGS entropy. It is remarkable that a  similar property is proved in \cite{LYZ0} 
for the generalized Gaussians and \ $\mathcal{N}_\lambda$. \ The exact statement is as follows: Let \ $\rho$ \ be a probability density function in \ $\mathbb{R}$. \
If \ $p\in [0, +\infty]$, \ $a>\frac{1}{1+p}$ \ and both \ $\mathfrak{s}_p[\rho]$ \ and \ $\mathcal{N}_a [\rho]$ \ are finite, then 
\begin{equation}
                       \frac{\mathfrak{s}_p[\rho]}{\mathcal{N}_a[\rho]} \ \geq \ \frac{\mathfrak{s}_p[\mathcal{G}]}{\mathcal{N}_a[\mathcal{G}]}
\end{equation}
Equality in (43) holds if and only if \ $\rho (x) = G_t(x)$ \ for some \ $t\in (0,+\infty)$. \ For a generalization of this theorem, see \cite{LYZ5}. Inequality (43)
can also be interpreted as a Sobolev-type inequality, so it may not come as a surprise that further relations exist between (43) 
with the Sobolev, the log-Sobolev and Gagliardo-Nirenberg inequalities \cite{DD, CENV, Villani}. We have, however, been unable to find any use of these inequalities
to the case of $q$-entropy that may appear to have any physical significance, certainly not for the questions addressed in this work.\\

 Before closing this section, one may also wish to notice that the \ $p=1$ \ special case 
of the generalized Gaussians (30) is nothing else than the $q$-exponential functions which appear as the equilibrium distributions, upon maximizing the
$q$-entropy under the usually employed constraints reflecting the classical ensembles in Statistical Mechanics \cite{T-book}. Hence (43) can be interpreted as a formalization and extension
of the well-known statements about the role of $q$-exponentials in the part of ``nonextensive" Statistical Mechanics employing the $q$-entropy \cite{T-book}.  \\  


\section{Fisher information, the Cram\'{e}r-Rao inequality and generalizations}

The Fisher information, and its associated Riemannian metric, have been fundamental concepts in Statistics since \cite{Rao, Jeffreys}. 
If we consider a probability distribution \ $\rho(x; \xi)$ \ where \ $x\in\mathbb{R}$ 
for simplicity, and in order to continue the arguments given above,\ and \ $\xi\in\mathbb{R}$, \
then one defines the Fisher information as the expectation value \ $\mathbb{E}$  \ given by
\begin{equation} 
          \mathcal{F}(\xi) \ = \ \mathbb{E} \left[  \left( \frac{\partial \log\rho (x; \xi)}{\partial \xi}\right)^2   \right]          
\end{equation}
or, in other words,
\begin{equation}
          \mathcal{F}(\xi) \ = \  \int_\mathbb{R} \left(\frac{\partial \log\rho(x;\xi)}{\partial \xi}\right)^2 \rho(x; \xi) \ dx 
\end{equation}
The multi-dimensional parameter space generalization of (44), (45) for \ $\xi\in \mathbb{R}^n$, \ or more generally for \ $\xi = (\xi^1, \ldots, \xi^n) \in\mathcal{M}$, \ 
where \ $\mathcal{M}$ \  is assumed to be the parameter space which is  an $n$-dimensional Riemannian manifold is
\begin{equation}
         \mathcal{F}_{ij}(\xi) \ = \ \mathbb{E} \left[\frac{\partial\log\rho(x;\xi)}{\partial \xi^i} \cdot \frac{\partial\log\rho(x;\xi)}{\partial \xi^j}\right], \hspace{10mm} i,j = 1, \ldots, n
\end{equation}  
or, in other words,
\begin{equation}
       \mathcal{F}_{ij}(\xi) \ = \ \int_\mathbb{R} \frac{\partial\log\rho(x;\xi)}{\partial \xi^i} \cdot  \frac{\partial\log\rho(x;\xi)}{\partial \xi^j} \ \rho(x;\xi) dx
\end{equation}
or after integration by parts, one gets the quadratic form on \ $\mathcal{M}$ 
\begin{equation}
         \mathcal{F}_{ij}(\xi) \ = \ \mathbb{E} \left[ - \frac{\partial^2\log\rho(x;\xi)}{\partial \xi^i\partial \xi^j}  \right]   
\end{equation}
We can also re-express (46) as
\begin{equation}
        \mathcal{F}_{ij}(\xi) \ = \ 4 \int_\mathbb{R} \frac{\partial\sqrt{\rho(x;\xi)}}{\partial\xi^i} \cdot \frac{\partial\sqrt{\rho(x;\xi)}}{\partial\xi^j} \ \rho(x;\xi) dx     
\end{equation}

As for any mathematical concepts, one can wonder what would their 
significance be, if any, for Statistical Mechanics in general, or for the questions addressed in this work in particular. We will not comment about the former question.
However, for the case of determining a relative $q$-entropy, and its possible dynamical underpinnings expressed by coarse-graining through convex bodies in the
phase space of a system of many degrees of freedom described by a Hamiltonian, someone can say the following. \\

First, one can see that the Fisher information (44), (45), 
or the Fisher metric in general (46), (47), is the Hessian (48) of the Kullback-Leibler divergence (4). Moreover, due to (49), it is a positive semi-definite quadratic form and as such, 
it can be seen as  providing a  
Riemannian metric on the parameter space \ $\mathcal{M}$. \ 
This is the fundamental point of a long line of investigations coming under the title ``Information Geometry"
whose scope far supersedes its potential applicability in Statistical Mechanics \cite{Amari}. The viewpoint of these investigations may prove to be important though for $q$-entropy purposes. 
One can see the Fisher information, or the Fisher metric, as analytical and geometric structures on the space of probability measures of the parameter space \ $\mathcal{M}$. \ A similar
space, the Monge-Kantorovich-Rubinstein-Vasherstein, or in the more widely used terminology the Wasserstein space, has attracted considerable attention recently
\cite{Villani, Ambrosio}. We claimed in \cite{NK9} that the Wasserstein space may be important for exploring the dynamical foundations of the $q$-entropy. So, in a sense, 
an investigation about the foundations of the $q$-entropy, via the Fidsher metric or via any other means, also provides a dynamical foundation for the parts of Information Geometry 
that may be related to an underlying Hamiltonian evolution of a system of many degrees of freedom. \\
      
 Second, according to Chentsov's theorem \cite{Cen1, Cen2}, the Fisher information metric (46) is the unique Riemannian metric on the parameter space \ $\mathcal{M}$ \ which is invariant under
sufficient statistics. The theorem \cite{Cen1, Cen2} is applicable for finite sample spaces. For infinite sample spaces see \cite{AJLS}.
Sufficient statistics is a stronger requirement than mere reparametrization invariance, which all geometric structures should obey. Without going into any details, 
as they can be found in the literature \cite{Amari, KV} and are not required in the sequel, sufficient statistics refers to mapping between sample spaces that preserves all information about a 
random variable. The Fisher metric can be straightforwardly checked to be invariant under sufficient statistics. However, it is much harder to prove that it is the only invariant under 
sufficient statistics.  One can hardly over-emphasize the use of Riemannian metrics in modelling physical systems, especially if such metrics have desirable features rendering them unique 
for modelling such systems.\\      

One sees  that the functional form of the Fisher information (44) involves a logarithm, hence that it may be somehow related to the BGS entropy or the Kullback-Leibler divergence in 
their Information theoretical applications. A question that can be posed is whether it is possible to extend the definition of the Fisher information to an analogous  quantity more closely 
related to the R\'{e}nyi-/$q$-entropy functionals (17), (21). A proposal for such generalized Fisher information was also provided in \cite{LYZ0}. 
Let \ $p\in [1, +\infty]$ \ and \ $a\in\mathbb{R}$. \
Then define the $(p, a)$-th Fisher information \ $\mathcal{F}_{p,a}[\rho]$ \ of a probability density \ $\rho$ \   as follows: For \ $p\in(1, +\infty)$, \ let \ $p^\ast \in (1, +\infty)$ \
denote its harmonic/convex conjugate, namely
\begin{equation}   
               \frac{1}{p} + \frac{1}{p^\ast} \ = \ 1
\end{equation}
Define
\begin{equation}
         \mathcal{F}_{p, a}[\rho] \ = \ \left\{ \int_\mathbb{R} \Big| [\rho(x)]^{a-2} \ \frac{d\rho(x)}{dx}\Big|^{p^\ast} \rho(x) \ dx  \right\} ^\frac{1}{p^\ast a}
\end{equation}
as long as the absolute value above is finite. For \ $p=1$ \ 
\begin{equation}  
                      \mathcal{F}_{p,a} [\rho] \ = \ \mathrm{ess} \sup \left\{ \Big| [\rho(x)]^{a-2} \  \frac{d\rho(x)}{dx}\Big|, \ \ x\in \mathrm{supp} \rho \subset \mathbb{R}    \right\}
\end{equation}
where \ $\mathrm{supp}$ \ stands for ``support of", under the assumption that \ $\rho$ \ is absolutely continuous, and that the essential supremum in (52) is finite. For the case
\ $p=+\infty $ \ one defines 
\begin{equation} 
            \mathcal{F}_{p,a}[\rho] \ = \ \inf_{K\in P(\mathbb{R})}    \left\{ \sum_{k\in K} \Bigg| \frac{[\rho(x_k)]^a}{a} - \frac{[\rho(x_{k-1})]^a}{a} \Bigg|  \right\}    
\end{equation}
where the index set \ $K$ \ indicates a partition of \ $\mathbb{R}$ \ and \ $P(\mathbb{R})$ \ indicates all possible partitions of \ $\mathbb{R}$. \ Definition (53) expresses the total variation of \ 
$\rho^a / a$, \ so we assume that \ $\rho^a$ \ has bounded variation in order for it  to make sense.\\  

The calculation of the $(p,a)$-th Fisher information for the generalized Gaussians (30), (32), (33)  is straightforward and gives, for \ $p\in[1, +\infty]$ \ and \ $a>\frac{1}{1+p}$ \ 
\begin{equation}
   \mathcal{F}_{p,a} [\mathcal{G}] \ = \ \left\{
                                \begin{array}{ll}
                                      c_{p,a}^\frac{a-1}{a} p^\frac{1}{a} \left(a-1+ap \right)^{-\frac{p-1}{pa}} & \mathrm{if} \ \ p < +\infty\\
                                                                                                   &                           \\
                                      \frac{2^\frac{1-a}{a}}{a^\frac{1}{a}} &    \mathrm{if} \ \ p=+\infty
                                \end{array}
                          \right.
\end{equation}
It may worth observing at this point that
\begin{equation}
             \mathcal{N}_a [\mathcal{G}] \ = \ a \ \mathfrak{s}_p[\mathcal{G}] \  (\mathcal{F}_{p,a}[\mathcal{G}])^a
\end{equation}
Stam's inequality \cite{Stam} provides an alternative characterization of the Gaussians as extremizing distributions. It states that among all probability distributions 
with the same Fisher information, Gaussians are the unique distributions that minimize the BGS entropy. Stam's theorem was generalized for the $(p,a)$-th Fisher information, the 
absolute entropy \ $\mathcal{N}_a$, \ and the generalized Gaussians in \cite{LYZ0} as follows. Let \ $p\in [1, +\infty]$, \ $a > \frac{1}{1+p}$ \ and let \ $\rho$ \ be a probability distribution on 
\ $\mathbb{R}$. \ For $p$: finite, \ $\rho$ \ is assumed to be absolutely continuous, as in the definition of the $(p,a)$-th Fisher information. Analogously, for \ $p=+\infty$, \
$\rho^a$ \ is assumed to have bounded variation. If  both \ $\mathcal{N}_a[\rho]$ \ and \ $ \mathcal{F}_{p,a}[\rho]$ \ are finite, then 
\begin{equation}        
                  \mathcal{N}_a[\rho] \ \mathcal{F}_{p,a}[\rho] \ \geq \ \mathcal{N}_a [\mathcal{G}] \ \mathcal{F}_{p,a} [\mathcal{G}]
\end{equation}
Equality holds in (56) if and only if there exists a \ $t>0$ \ and an \ $x_0\in\mathbb{R}$ \ such that \ $\rho(x) = \mathcal{G}_t (x-x_0)$, \ for all $x\in\mathbb{R}$.  
This is another extremizing characterization of the generalized Gaussians, after (43). \\

An important relation involving the Fisher information is the Cram\'{e}r-Rao inequality \cite{CT}, a fundamental inequality in statistical inference. Given an unbiased estimator, 
the Cram\'{e}r-Rao inequality provides a lower bound for its variance in terms of the Fisher information in the case of a scalar/single parameter. It provides a lower bound about the 
estimator's covariance in terms of the the Fisher metric (matrix) in the multi-dimensional parametric/vector case.  Without providing too many details, 
as they can be readily found in the literature \cite{CT, KV}, consider a probability distribution \ $\rho(x;\xi)$ \ which depends on a single parameter $\xi$.
Then the variance of an unbiased estimator $\hat{\Xi}$  of $\xi$ obeys the Cram\'{e}r-Rao inequality       
\begin{equation}
                 \mathrm{var}(\hat{\Xi}) \ \geq \ \frac{1}{\mathcal{F}(\xi)} 
\end{equation}
where \ $\mathcal{F}(\xi)$ \ stands for the Fisher information (44). 
If a \ $\hat{\Xi}$ \ is an unbiased estimator of a vector-valued parameter $\xi = (\xi^1, \ldots, \xi^n)$, \ $\xi\in\mathbb{R}^n$
the Cram\'{e}r-Rao inequality generalizes as 
\begin{equation}
             \mathrm{cov}(\hat{\Xi}) \ \geq \ \mathcal{F}^{-1} 
\end{equation}   
where \ $\mathrm{cov}$ \ stands for the covariant matrix of the estimator \ $\hat{\Xi}$, \ and \ $\mathcal{F}$ \ is the matrix form of the Fisher metric (46). 
The generalization to the case of the $(p,a)$-th Fisher information (51), (52), (53) and the generalized Gaussians goes as follows \cite{LYZ0}: 
Let \ $p\in[1, +\infty]$, \ $a > \frac{1}{1+p}$. \  If both \ 
$\mathfrak{s}_p[\rho]$ \ and \ $ \mathcal{F}_{p,a}[\rho]$ \ are finite for a probability density \ $\rho$, \ then 
\begin{equation}   
                   \mathfrak{s}_p [\rho] \ \mathcal{F}_{p,a} [\rho] \ \geq \ \mathfrak{s}_p[\mathcal{G}]  \ \mathcal{F}_{p,a}[\mathcal{G}]
\end{equation}
The equality holds in (59) if and only if \ $\rho = \mathcal{G}_t$, \ for some \ $t>0$. \ In (59), if $p$: finite, then \ $\rho$ \ is assumed to be absolutely continuous, and if \ $p=+\infty$ \
then \ $\rho^a$ \ is assumed to have bounded variation. This is a third extremal characterization of the generalized Gaussians, after (43), (56).\\

We see that the definitions of the relative divergence (17), the absolute entropy (36)  and the corresponding generalized Gaussians (30), (32), (34), (35) give us generalizations 
of classical inequalities that are well-known in Information Theory. Definitions such as (17) and subsequently (20), (21) may be reasonable, and in particular (21)
may be a good guess for an expression of the relative $q$-entropy. The relations with convex geometry of these functionals, which we did not elaborate upon in this work,
but can be found in the references, are an element which increases our confidence toward the relevance of these expressions for Physics. 
Ultimately, only the calculations of such functionals in particular physical models and quantities stemming from them, and how they compare with experimental data will 
be the judge of the usefulness, if any, of such expressions in  Physics.\\         


\section{Conclusions and outlook}

In this work, we presented a proposal about the ``essential" functional form (17) of a relative $q$-entropy (21), (22) a functional form which is also shared, up to a power and a logarithm, 
with a variation of the relative R\'{e}nyi entropy (20). This form appears in the fundamental work of Lutwak-Yang-Zhang, which relies on ideas of the $L_p-$ Brunn-Minkowski theory and its dual.
We also stated the extremal distributions of this functional form, an $L_p$ form of the Fisher information, and the related form of a 
generalized Cram\'{e}r-Rao inequality. We attempted to point out the significance of these constructions, as judged from the viewpoint of the $q$- entropy and their possible significance 
for the part of Statistical Mechanics based on the $q$-entropy. 
Perhaps mirroring the precendence of the construction of the $q$-entropy in Information Theory, before its re-discovery by C. Tsallis in \cite{T1}, and its further development in 
some parts of Physics  during the last three decades, the relative $q$-entropy functional (21), (22)  may prove to be of interest in Statistical Mechanics and Complexity Theory in the future.\\  

The present work takes a considerably different path from the recent \cite{BCP, RPL, Rast, ZSLL, Italians, ZY, WLi, ShiHan} which refer mostly to relative $q$-entropies for quantum systems, 
in an attempt to quantify coherence measures. The difference is not just superficial: indeed one could attempt to extend the LYZ functional to Quantum Physics in the usual way, by replacing
the probability distributions by appropriately defined pseudo-differential operators acting on  Hilbert spaces of states of such quantum systems. Whether such a naive substitution works
is a totally different matter altogether as one faces, once more, the notorious ``operator ordering problem" which plagues all attempts to quantization starting from classical models, 
but is especially acute in attempting to quantize General Relativity in a background-independent and non-perturbative manner.\\    
  
The dual $L_p-$ Brunn-Minkowski theory and the relations with the better known $L_p-$ Brunn-Minkowski theory, have the potential of 
providing a partial  understanding of the dynamical foundations of the $q$-entropy, at least as they pertain to coarse-graining 
\cite{DeGos1, DeGos2, DeGos3, DeGos4, NK10, Gos0}. The idea of such an application of the Brunn-Minkowski theory to entropy functionals is not really new. 
However, this geometric viewpoint has not been advocated, much less appreciated, in Physics yet, so far as we know. 
Elements of such a viewpoint  can be traced to the relations between Information Theory and Brunn-Minkowski theory as can be seen in  
\cite{CC, DCT, CT}, for instance.\\   

Based, in part on the present work, one could conjecture that coarse graining with convex polyhedra in an $L_p$ sense in phase space,  
is a reason for choosing to describe the collective behavior of a Hamiltonian system of many degrees of freedom by using the $q$-entropy. 
Dually, one could employ in such a coarse-grained description star-shaped 
bodies in an $L_p$ sense in phase space, as the latter are the fundamental objects of the dual $L_p-$ Brunn-Minkowski theory. 
Such coarse-graining procedures have, ideally, to be somehow related to the dynamical foundations of the theory. 
Whether they may have anything to do with Physics, can ultimately only be decided by comparing its implications with the results of experiments.\\

One step in this general direction of coarse-graining was taken by the  introduction of 
``quantum blobs"  \cite{Gos1} which are phase space volumes  invariant only under linear, rather than the fully nonlinear, symplectic maps, of the phase space.
The proposal for the  existence and use of ``quantum blobs" is firmly rooted on the symplectic non-squeezing theorem \cite{Gr} and, more generally, on the existence 
and properties of symplectic capacities \cite{DS, GosL}. These results provide some genuinely 2-dimensional restrictions to deformations, under symplectic maps, of 
phase space volumes. This rigidity of symplectic maps however, is  only applicable to 
projections of phase space volumes on symplectic 2-planes of the phase space and does not hold for sections of phase space volumes \cite{AM}. 
It may be worth mentioning at this point, that the implications for Statistical Mechanics, if any, of the distinction between symplectic and volume-preserving maps 
is currently largely unknown \cite{Gos2, NK12, NK13}. Addressing this question  may have far-reaching consequences for a better understanding of the foundations 
of Statistical Mechanics especially as it applies to ``complex systems" or to systems out of equilibrium.\\         

 In the same isoperimetric/Sobolev inequality extremal spirit presented in this work, recall that the 
Viterbo conjecture \cite{Vit, AAKO} essentially claims that the Euclidean ball has the maximum symplectic capacity among all convex sets in the standard symplectic space\ $\mathbb{R}^n$ \ 
with a given volume, for all symplectic capacities, is true for $L_p$ balls. However, notice that it is violated for convex sets which are sections of star-shaped bodies. 
Hence sections and projections seem to behave quite differently from a symplectic viewpoint, unlike their familiar correspondence via polar duality encountered in Functional Analysis and in Convex Geometry.  \\
 
 In its most straightforward interpretation, the duality of the Brunn-Minkowski with the dual Brunn-Minkowski theories rests in the replacement of concepts involving projections to 
concepts involving sections of  convex and of star-shaped bodies. Hence a naive use of the ``quantum blobs" or their nonlinear symplectic analogues, if such structures could be reasonably defined, 
may not be the most appropriate objects for providing a form of coarse-graining of phase space, which may behave well under the duality between the $L_p-$ Brunn-Minkowski and 
the dual $L_p-$ Brunn-Minkowski theories.  Such a duality, if it exists, may be used in establishing and explaining a suspected invariance of the $q$-entropy under what appears to be a set 
of M\"{o}bius-like transformations of the nonextensive parameter $q$ \cite{NK11}. \\

It should also be noticed that ``quantum blobs" are essentially Riemannian constructions \cite{Gos1}. As such, they need to be generalized in order 
to incorporate the $q$-exponentials and generalized $L_p$ Gaussians which arise as extremal distributions of  the above proposed relative $q$-entropy functionals (17), (20), (21). 
It is, however, not clear to us how to define such essentially $L_p$ generalizations of the
``quantum blobs", if possible at all. Moreover, we are not certain which invariance requirements someone would have to impose to derive such structures, 
and how such requirements could be justified on dynamical grounds for, at least, the Hamiltonian systems of many degrees of freedom which are relevant to Statistical Mechanics.  
We believe that it may be worth pursuing and further developing some of these ideas in the future.\\ 

                         \vspace{5mm}


\noindent{\bf Acknowledgements:} \ We are grateful to the referees whose constructive criticism helped improve the clarity of the exposition of this work.\\


              \vspace{3mm}



\end{document}